\documentclass[cameraready]{Interspeech}
\usepackage{flushend}
\usepackage[all]{nowidow}

\title{SGPA: Spectrogram-Guided Phonetic Alignment for Feasible Shapley Value Explanations in Multimodal Large Language Models}

\author[affiliation={1}, orcid=0009-0007-0728-1043, equalcontribution, correspondingauthor]{Paweł Dominik}{Pozorski}
\author[affiliation={1}, orcid=0009-0000-2797-6044, equalcontribution, correspondingauthor]{Jakub Miłosz}{Muszyński}
\author[affiliation={1}, orcid=0000-0001-7714-4844]{Maria}{Ganzha}

\address{
    $^1$ Warsaw University of Technology, Warsaw, Poland
}

\email{\texttt{\{jakub.muszynski2.stud, pawel.pozorski.stud, maria.ganzha\}@pw.edu.pl}}


\usepackage{comment}

\begin{document}

\maketitle

\begin{abstract}
    Explaining the behavior of end-to-end audio language models via Shapley value attribution is intractable under native tokenization: a typical utterance yields over $150$ encoder frames, inflating the coalition space by roughly $10^{42}$ relative to text; individual audio frames lack standalone meaning; and token boundaries that bisect phonetic transitions introduce masking artifacts. We introduce Spectrogram-Guided Phonetic Alignment (SGPA), a four-stage pipeline that combines Connectionist Temporal Classification forced alignment with spectral boundary refinement to produce acoustically stable, word-aligned audio segments. Controlled diagnostics on LFM2-Audio-1.5B with VoiceBench show that SGPA yields a 43$\times$ reduction in model evaluations. Statistical testing confirms that SGPA significantly alters attribution concentration while preserving the global cumulative profile, establishing it as a feasibility-enabling layer for audio explainability.
\end{abstract}

\section{Introduction}

The rapid evolution of Multimodal Large Language Models (MLLMs) has produced systems that process audio end-to-end, bypassing traditional cascaded speech-to-text pipelines~\cite{yang2025largelanguagemodelsmeet, Lee_2022}. While these architectures achieve strong performance, their internal decision-making remains opaque: when an audio input drives a particular response, there is no principled way to determine which parts of the signal were most influential~\cite{zhao2023explainabilitylargelanguagemodels}. Shapley value (SV) attribution~\cite{RM-670-PR}, popularized in machine learning by Lundberg and Lee~\cite{lundberg2017unifiedapproachinterpretingmodel} and recently extended to token-level LLM explanations by Horovicz and Goldshmidt~\cite{goldshmidt2024tokenshapinterpretinglargelanguage}, offers a theoretically grounded framework for this analysis. However, applying SV to audio inputs introduces barriers that do not arise for text~\cite{arrieta2020explainable}.\penalty-10000

The root cause is native audio tokenization. Unlike text, which discretizes naturally into words or sub-words, audio is processed as a sequence of dense encoder frames~\cite{yang2025largelanguagemodelsmeet}. This representation introduces three critical barriers for SV-based audio analysis:

\begin{itemize}
    \item \textbf{Dimensionality explosion.} A standard $3$-second utterance produces over $150$ encoder frames under native tokenization. Since exact SV computation requires evaluating all $2^n$ subsets, this inflates the coalition space from $2^{10}$ (for a 10-word text) to $2^{150}$ -- a factor of ${\approx}10^{42}$. For a representative $50$-token prompt, exact computation would require ${\approx}10^{15}$ model evaluations; at \SI{1}{\milli\second} per call this exceeds $30$~years.\footnote{Derived from $2^{50} \times 10^{-3}\,\text{s} \approx 3.6 \times 10^{10}$\,s\,$\approx 35$~years.}
    \item \textbf{Semantic dilution.} Individual \SI{20}{\milli\second} audio frames lack standalone meaning. Attributions assigned to these opaque units appear as granular noise that cannot be interpreted by human supervisors or linguistic experts~\cite{mosca-etal-2022-shap}.
    \item \textbf{Boundary artifacts.} Native token boundaries frequently bisect phonetic transitions such as co-articulations~\cite{ladefoged2014course, hardcastle1999coarticulation}. Masking tokens at these positions introduces audible discontinuities and out-of-distribution perturbations that degrade the fidelity of SV estimates.
\end{itemize}

This work addresses these challenges by introducing \textbf{Spectrogram-Guided Phonetic Alignment (SGPA)}, a four-stage preprocessing pipeline that maps raw audio waveforms to word-aligned segments using Connectionist Temporal Classification (CTC)~\cite{graves2006connectionist} and local spectral refinement~\cite{mcfee2015librosa}. By replacing architecture-specific encoder frames with semantically grounded word-level segments, SGPA reduces the effective player count by $10$--$50\times$, compressing the required model evaluations from ${\approx}2{,}552$ to ${\approx}59$ per sample -- making SV-based audio attribution computationally feasible on consumer hardware.\penalty-10000

Our contributions are threefold: (i)~we present SGPA, a model-agnostic alignment layer for audio SV computation; (ii)~we provide controlled diagnostics demonstrating that SGPA is not a neutral transformation -- it significantly alters attribution concentration (Cohen's~$d$~\cite{cohen1988statistical} up to \textit{$-1.37$}) while preserving the global cumulative SV profile; and (iii)~we release an open-source Python package, \texttt{mllm-shap}\footnote{\url{https://pypi.org/project/mllm-shap/}}, implementing the full pipeline. SGPA is integrated into a broader explainability platform available in the accompanying package~\cite{pozorskimuszynskimllmshap}.

Section~2 details the SGPA pipeline. Section~3 presents diagnostic experiments, and Section~4 discusses limitations and future directions.

\section{Spectrogram-Guided Phonetic Alignment}
\label{sec:sgpa}

\begin{figure}[t]
  \centering
  \includegraphics[width=\linewidth]{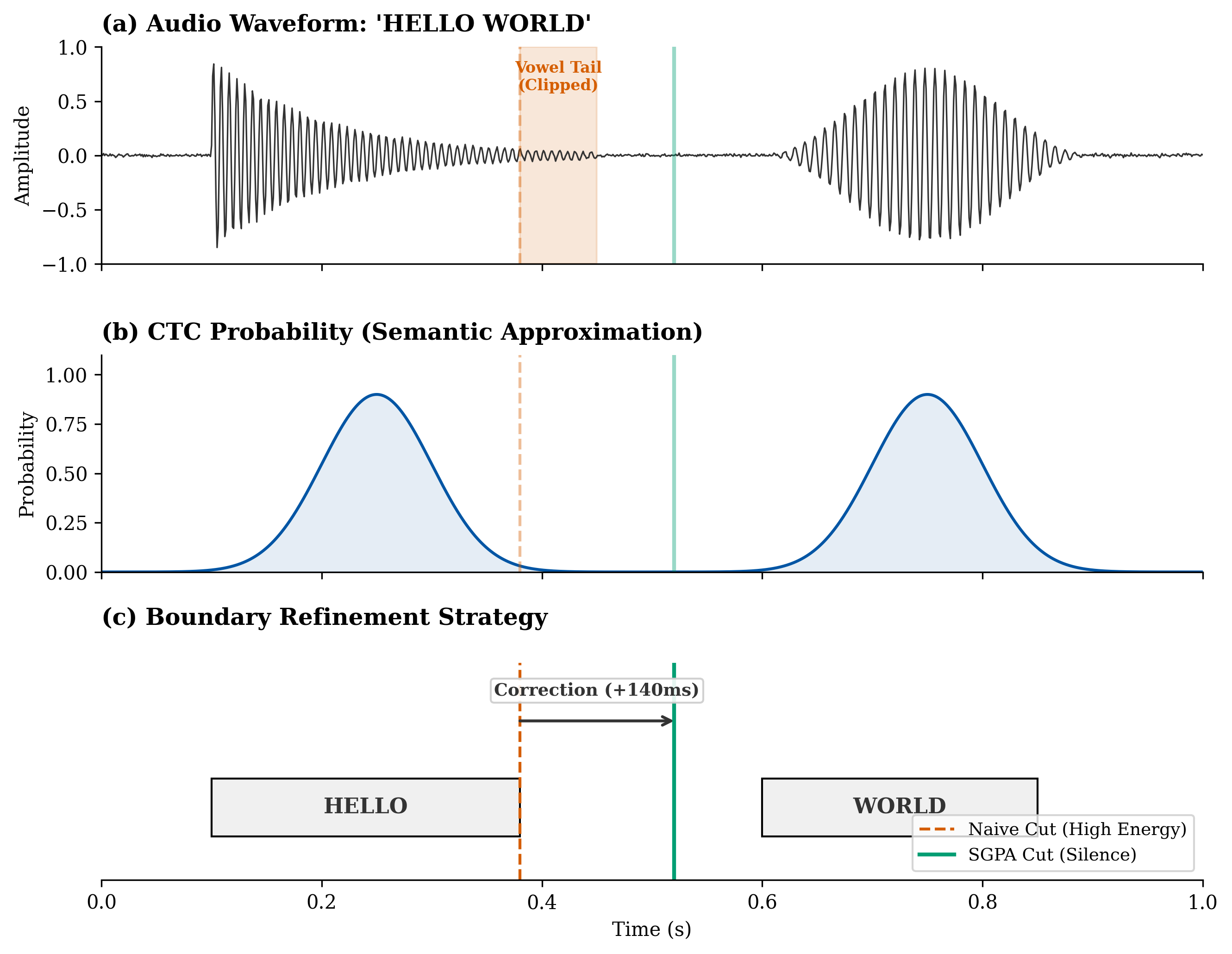}
  \caption{SGPA pipeline: transcript decomposition $\to$ CTC alignment $\to$ spectral boundary refinement $\to$ word-level aggregation. The CTC boundary (dashed) is shifted to a spectrally stable region (solid).}
  \label{fig:sgpa_pipeline}
\end{figure}

SGPA is a four-stage pipeline that converts a raw audio waveform and its transcript into word-aligned audio segments suitable for SV analysis. The key idea is to redefine SV \emph{players}: instead of opaque encoder frames, each word in the transcript becomes a single player whose audio boundaries are determined by forced alignment and local spectral cues. When a player is excluded from a coalition, its corresponding audio segment is replaced with silence (zero amplitude)~\cite{covert2021explaining}, preserving the temporal structure of the remaining signal. Figure~\ref{fig:sgpa_pipeline} provides an overview.

\subsection{Stage~1: Transcript Decomposition}

The input transcript is decomposed into words and characters, recording a character-to-word membership map. This bookkeeping is essential because Stage~2 operates at the character level, while the final SV game requires word-level segments. The membership map allows character-level time boundaries to be merged upward in Stage~4.

\subsection{Stage~2: Initial Alignment via CTC}

We feed the audio waveform into Wav2Vec2-XLSR-53~\cite{baevski2020wav2vec}, a self-supervised speech model chosen for its robust CTC alignment quality and public availability; its multilingual capability additionally enables future cross-lingual extension. The model produces an emission matrix $\mathbf{E} \in \mathbb{R}^{T \times V}$, where $T$ is the number of time frames and $V$ the vocabulary size. We apply Viterbi decoding~\cite{viterbi1967error} to extract the most likely character path:
\begin{align}
  \pi^* = \arg\max_{\pi}\, P(\pi \mid \mathbf{E})
  \label{eq:viterbi}
\end{align}
This yields approximate start and end frame indices for each character, converted to timestamps via the model's effective frame stride. These timestamps serve as coarse anchors but are not guaranteed to coincide with acoustically stable cut points.

\subsection{Stage~3: Spectral Boundary Refinement}

CTC-derived boundaries may fall in the middle of phonetic transitions. Speech commonly alternates between high-energy steady regions (e.g., vowels) and lower-energy regions or rapid transitions (e.g., closures, silences). For each candidate boundary~$t_{\mathrm{est}}$, we define a local search window $W = [t_{\mathrm{est}} - \delta,\, t_{\mathrm{est}} + \delta]$, where $\delta$ is a tunable half-width, and compute two signal-level features using \texttt{librosa}~\cite{mcfee2015librosa}:
\begin{itemize}
    \item \textbf{Short-time energy} $E[n]$: the RMS energy per frame, capturing local amplitude.
    \item \textbf{Spectral flux} $SF[n]$: the frame-to-frame $L^2$ change in magnitude spectrum, derived from the Short-Time Fourier Transform.
\end{itemize}
The refined boundary is selected as the point of minimal combined activity:
\begin{align}
  t_{\mathrm{final}} = \arg\min_{t \in W}\left(\alpha\, E[t] + \beta\, SF[t]\right)
  \label{eq:boundary}
\end{align}
where $\alpha{=}0.8$ and $\beta{=}0.2$ are empirically tuned weights that favor low-energy, spectrally stable regions -- typically inter-word pauses -- thereby reducing the risk of introducing audible artifacts during the masking operations required for SV estimation.

\subsection{Stage~4: Word-Level Aggregation}

The refined character-level boundaries are merged into word-level segments using the membership map from Stage~1. This granularity aligns with the primary unit of human interpretation (words), reduces the player count from $O(100)$ native tokens to $O(10)$ segments, and makes SV estimation with the Neyman-allocated complementary-contributions estimator~\cite{shapleyapproximations, nayman} computationally feasible.

\section{Diagnostic Evaluation}
\label{sec:experiments}

To validate SGPA, we conduct a controlled comparison between two segmentation regimes: (i)~the model's native audio tokenization and (ii)~SGPA word-aligned segments. All remaining estimator and inference settings are held constant. Reproduction notebooks are available in the project repository\footnote{\url{https://github.com/Pawlo77/MLLM-Shap}}.

\subsection{Setup}

\subsubsection{Model} We use \textbf{LFM2-Audio-1.5B}~\cite{LIQUID_LFM2_AUDIO_2025}\footnote{\url{https://huggingface.co/LiquidAI/LFM2-Audio-1.5B}}, a 1.5\,B-parameter open-source, end-to-end audio foundation model. Our selection criteria were: (i)~direct audio token processing without a cascaded speech$\,\to\,$text pipeline~\cite{yang2025largelanguagemodelsmeet, Lee_2022}; (ii)~publicly available weights; and (iii)~a parameter count small enough to fit within the \SI{16}{\giga\byte} VRAM of a consumer-grade Nvidia RTX~4080. LFM2 was the smallest publicly available model satisfying all three criteria.

\subsubsection{Dataset} We evaluate on a single-sentence English subset of \textbf{VoiceBench}~\cite{chen2024voicebench}, a benchmark that assesses voice assistants across general knowledge, instruction-following, and safety dimensions. To construct the subset, we: (i)~remove exact-match text duplicates across all VoiceBench partitions; (ii)~retain only single-sentence prompts via NLTK~\cite{bird2009natural} sentence segmentation; and (iii)~apply stratified sampling over the source sub-datasets, reducing the pool by $90\%$ to arrive at $100$ samples (mean $7.19$ tokens, $37.87$ characters). Audio is synthesized with Google Chirp~3~HD TTS\footnote{\url{https://cloud.google.com/text-to-speech}} using default settings, yielding mean durations of \SI{2.69}{\second} (male) and \SI{2.48}{\second} (female), with a combined corpus length of \SI{1035.5}{\second}. The dataset is publicly available\footnote{\url{https://huggingface.co/datasets/Pawlo77/mllm-shap}}.\penalty-10000

\subsubsection{Modes} Since SGPA targets audio inputs, we evaluate two speech-to-speech configurations: male voice (\texttt{SM2S}) and female voice (\texttt{SF2S}). Each prompt is synthesized in both variants to assess SGPA's consistency across speaker characteristics.

\subsubsection{SV approximation} We employ the Neyman-allocated complementary-contributions estimator~\cite{shapleyapproximations}, which reuses each coalition's complementary contribution across all players it contains, combined with the Neyman stratified sampling formula~\cite{nayman} for optimal budget allocation. The total budget is set to $3n^2$ model evaluations, where $n$ is the number of players. The algorithm proceeds in two phases: Phase~1 draws $m_{\mathrm{init}} = \max(2, \lfloor m / 2n^2 \rfloor)$ samples per stratum to initialize variance estimates; Phase~2 allocates the remaining budget proportionally to the estimated standard deviations, minimizing the total SV variance.

\subsubsection{Infrastructure} All experiments run on a single Nvidia RTX~4080 GPU (\SI{16}{\giga\byte} VRAM). The \texttt{mllm-shap} package\footnote{\url{https://pypi.org/project/mllm-shap/}} \cite{pozorskimuszynskimllmshap} implements SGPA, the SV estimators, and model connectors.

\subsection{Feasibility: Coalition Space Reduction}
\label{subsec:feasibility}

\begin{table}[t]
  \caption{Effect of SGPA on Neyman SV estimation cost. Mean model calls and wall-clock time per sample over $100$ VoiceBench single-sentence prompts.}
  \label{tab:sgpa_cost}
  \centering
  \begin{tabular}{llcc}
    \toprule
    \textbf{SGPA} & \textbf{Mode} & \textbf{Calls (mean)} & \textbf{Time (s)} \\
    \midrule
    \checkmark & SM2S & 59.42  & 67.53 \\
    \checkmark & SF2S & 59.32  & 66.08 \\
     --         & SM2S & 2365.76 & 1696.49 \\
     --         & SF2S & 2552.14 & 1819.88 \\
    \bottomrule
  \end{tabular}
\end{table}

Table~\ref{tab:sgpa_cost} reports computational cost under both segmentation regimes. Without SGPA, native audio tokenization produces long feature sequences and the Neyman estimator requires ${\approx}2{,}552$ model calls per sample (mode~\texttt{SF2S}), with a mean wall-clock time of \SI{1820}{\second}. With SGPA, word-level aggregation compresses the player set to ${\approx}7$ segments, reducing calls to ${\approx}59$ per sample -- a $43\times$ reduction -- and wall-clock time to ${\approx}\SI{66}{\second}$ ($28\times$ faster).

\begin{figure}[t]
  \centering
  \includegraphics[width=\linewidth]{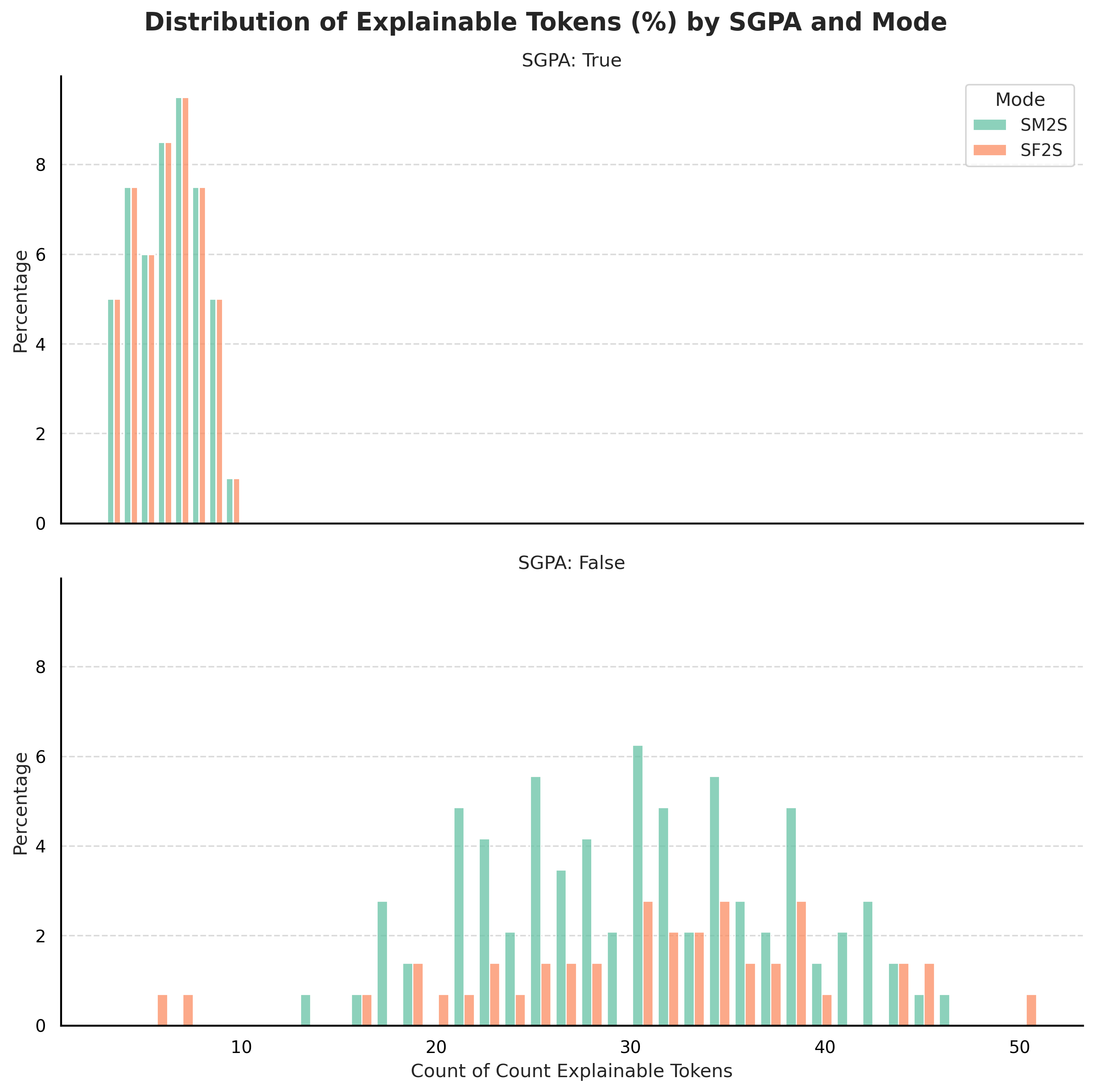}
  \caption{Distribution of explainable token counts per sample with and without SGPA across modes. SGPA concentrates counts in a narrow regime; native tokenization yields substantially longer and more dispersed sequences.}
  \label{fig:token_dist}
\end{figure}

Figure~\ref{fig:token_dist} illustrates the shift in feature-count distributions. With SGPA enabled, explainable lengths concentrate in a narrow, low-count regime (interquartile range $\approx$2); without SGPA, they expand substantially and become more dispersed, directly increasing coalition complexity.

The Neyman budget allocation also shifts qualitatively. With SGPA, the estimator distributes $37\%$ of its budget to the initial variance-estimation phase (Phase~1) and $63\%$ to the optimized phase (Phase~2). Without SGPA, the large game forces nearly all budget ($\geq98\%$) into Phase~2, leaving minimal room for reliable variance estimation and yielding a less stable approximation.

\subsection{Attribution Statistics}
\label{subsec:attribution}

Because SGPA redefines the player partition -- from model-native audio units to word-aligned segments -- it alters the cooperative game being solved. Attribution statistics under the two regimes are therefore expected to differ; the question is how and by how much. To quantify this, we compare three concentration metrics.

\subsubsection{Metrics} Let $\tilde{s}_i = |s_i| / \sum_j |s_j|$ denote the normalized absolute SV for player~$i$. We report:
\begin{itemize}
    \item \textbf{Top-$20\%$ mass}: $\sum_{i=1}^{\lceil 0.2n \rceil} \tilde{s}_{(i)}$, the share captured by the top quintile.
    \item \textbf{Gini coefficient}: $G = \frac{1}{2n}\sum_{i}\sum_{j} |\tilde{s}_i - \tilde{s}_j|$, measuring inequality.
    \item \textbf{Normalized entropy}: $H_{\text{norm}} = {-\sum_i \tilde{s}_i \log \tilde{s}_i}\, /\, {\sqrt{n}}$. Division by $\sqrt{n}$ compensates for the length dependence of raw entropy, enabling comparison between SGPA (${\approx}7$ players) and native tokenization (${\approx}50$ players).
\end{itemize}
\subsubsection{Results} We apply paired $t$-tests with Bonferroni correction ($\alpha = 0.05$). Table~\ref{tab:sgpa_stats} reports the results. Nearly all comparisons are significant, confirming that SGPA systematically alters attribution concentration.

\begin{table}[t]
  \caption{Paired significance tests of per-observation SV summary metrics (SGPA~\emph{on} vs.\  \emph{off}), Bonferroni-corrected $\alpha = 0.05$. Entropy is normalized by $\sqrt{n}$.}
  \label{tab:sgpa_stats}
  \centering
  \begin{tabular}{llccc}
    \toprule
    \textbf{Mode} & \textbf{Metric} & $p$ & \textbf{Cohen's $d$} & \textbf{Sig.} \\
    \midrule
    SM2S & Gini        & $<$0.01 & 0.50  & \checkmark \\
         & Entropy     & $<$0.01 & $-$1.37 & \checkmark \\
         & Top-20\%    & $<$0.01 & 0.86  & \checkmark \\
    \addlinespace
    SF2S & Gini        & 0.01    & 0.21  & --- \\
         & Entropy     & $<$0.01 & $-$0.97 & \checkmark \\
         & Top-20\%    & $<$0.01 & 0.72  & \checkmark \\
    \bottomrule
  \end{tabular}
\end{table}

The large negative Cohen's~$d$ for normalized entropy ($-1.37$ for \texttt{SM2S}, $-0.97$ for \texttt{SF2S}) indicates that SGPA \emph{increases} entropy, spreading attribution more evenly across the fewer word-level players. Conversely, the positive~$d$ for top-$20\%$ mass shows that each player's share grows as the total count shrinks -- a mechanical consequence of distributing the same total attribution over fewer units.

\begin{figure}[t]
  \centering
  \includegraphics[width=\linewidth]{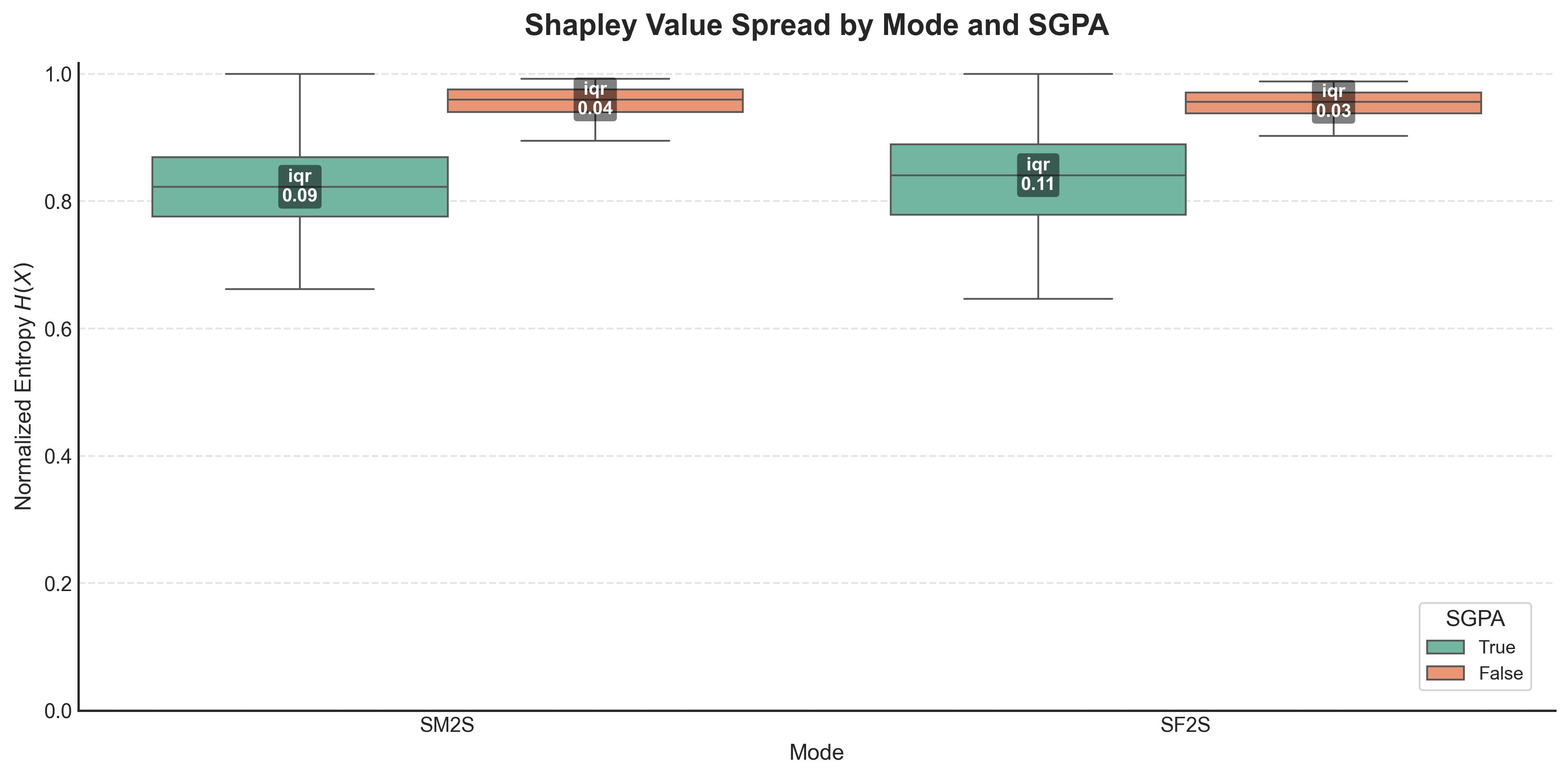}
  \caption{Normalized attribution entropy ($H / \sqrt{n}$) by mode with and without SGPA. SGPA widens the interquartile range, reflecting greater variability in concentration across samples.}
  \label{fig:entropy}
\end{figure}

The sole non-significant comparison (\texttt{SF2S}~Gini, $d{=}0.21$) suggests that overall inequality is less sensitive to segmentation for female-voice inputs. Figure~\ref{fig:entropy} visualizes this effect: SGPA expands the interquartile range of normalized entropy, indicating a broader set of concentration regimes across observations.

\subsection{Preservation of Macro-Trends}
\label{subsec:macro}

\begin{figure}[t]
  \centering
  \includegraphics[width=\linewidth]{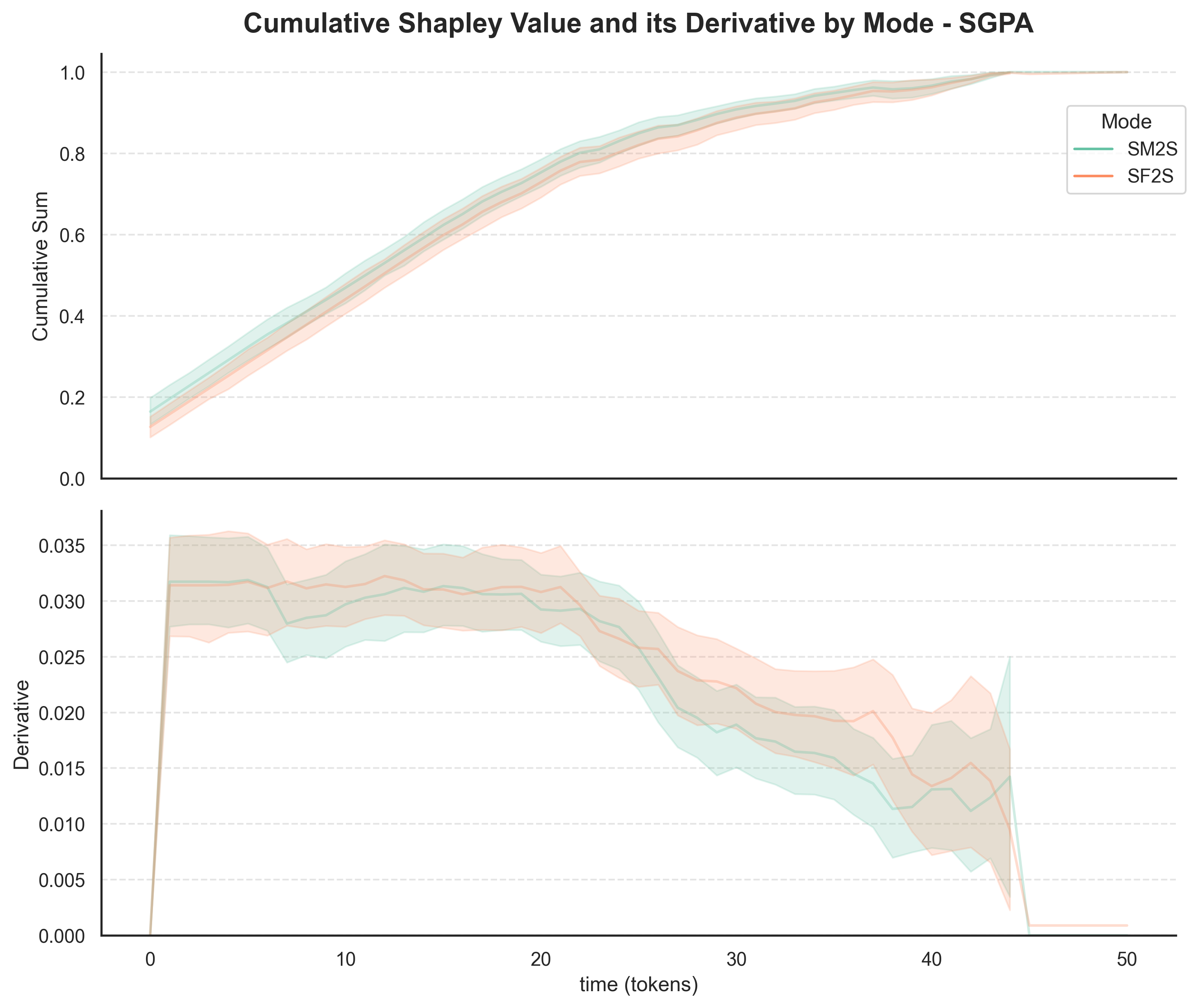}
  \caption{Position-normalized cumulative SV profile (with derivative) for speech-to-speech mode, illustrating the preserved ''early mass'' macro-trend under SGPA.}
  \label{fig:cumulative_sv}
\end{figure}

A key methodological question is whether SGPA alters the qualitative attribution trajectory. Figure~\ref{fig:cumulative_sv} addresses this by comparing position-normalized cumulative SV profiles. Both conditions exhibit an ''early mass'' pattern -- the first few positions accumulate a disproportionate share of total attribution -- followed by gradual decay. SGPA modifies local allocation density (the fine-scale derivative structure) while preserving the broader accumulation profile. This is expected: SGPA aggregates fine-grained audio units into semantically interpretable segments, which dampens high-frequency variations attributable to micro-boundary differences under native tokenization, while the global trajectory remains intact.

\section{Discussion and Conclusion}
\label{sec:conclusion}

We introduced SGPA, a four-stage alignment pipeline that converts raw audio waveforms into word-aligned segments for SV attribution. By combining CTC-based forced alignment with spectrogram-guided boundary refinement, SGPA addresses the three barriers that make SV computation on audio intractable under native tokenization. Our controlled diagnostics on LFM2-Audio-1.5B with VoiceBench yield three principal findings:

\begin{enumerate}
    \item \textbf{Feasibility.} SGPA compresses the player set from ${\approx}50$ native tokens to ${\approx}7$ word-aligned segments, reducing model evaluations by $43\times$ (from 2{,}552 to 59) and wall-clock time from ${\approx}30$~minutes to roughly one minute per sample on a consumer-grade GPU.
    \item \textbf{Non-neutrality.} SGPA fundamentally changes the cooperative game being solved: it increases normalized entropy (Cohen's $d = -1.37$ for \texttt{SM2S}) and raises top-quintile mass ($d = 0.86$), confirming that word-level segmentation redistributes attribution relative to native frames -- a deliberate trade-off favoring interpretability over fidelity to the model's internal tokenization.
    \item \textbf{Structural preservation.} Despite altering local allocation density, SGPA preserves the global cumulative SV profile and its characteristic ''early mass'' pattern, indicating that the word-level partition retains the model's broad attribution structure.
\end{enumerate}

\subsection{Limitations} Four limitations should be noted. First, SGPA's spectral refinement weights ($\alpha{=}0.8$, $\beta{=}0.2$) and the search-window half-width~$\delta$ were empirically tuned on English data; their robustness to tonal languages or phonologically diverse inventories has not been evaluated. Second, all diagnostics used a single model (LFM2-Audio-1.5B) due to research hardware constraints; replication on larger architectures such as Kimi Audio \cite{KimiTeam_2025} would strengthen generalizability. Third, because SGPA redefines the player partition, attributions under SGPA and native tokenization are not directly comparable -- this is an inherent property of SV, where explanations depend on the chosen player set~\cite{RM-670-PR}. Fourth, SGPA requires a transcript; audio-only applications would need upstream ASR, introducing an additional error source.

\subsection{Future work} Future work will extend validation to additional architectures, cross-lingual settings, and alternative SV estimators. The fixed spectral weighting could be replaced with learned boundary predictors. Finally, phrase-level and sub-word-level aggregation strategies may offer complementary granularity trade-offs for different analytical objectives.

\section{Generative AI Use Disclosure}
Generative AI tools were used for editing and polishing the manuscript. All content was reviewed and validated by the authors, who take full responsibility.

\bibliographystyle{IEEEtran}
\bibliography{mybib}

\end{document}